\documentclass[12pt]{article}
\usepackage{amsfonts}
\usepackage{latexsym}
\usepackage{amstext}
\usepackage{sc3conf1}

\newcommand{\beq}{\begin{equation}}
\newcommand{\eeq}{\end{equation}}
\newcommand{\beqa}{\begin{eqnarray}}
\newcommand{\eeqa}{\end{eqnarray}}
\newcommand{\nn}{\nonumber \\}

\newcommand{\Fraa}[2]{
#1 {}_/^{{}_{}^{}}{\!\! /} \!\!\!\!{}_{}^{}
\begin{array}{c} \, \\ #2 \end{array}\!\!\!}

\def \R {{\mathbb R}}

\def \Z {{\mathbb Z}}
\def \N {{\mathbb N}}

\def \Sr {{\mathbb S}}
\def \M {\overline{M}}
\def \W {\mathcal{W}}
\def \Wt {\mathcal{W}^{\, t}}

\def \lvac {\left\langle 0 \! \left| \right. \right.  \!}
\def \rvac {\!\! \left. \left. \right| \! 0 \right\rangle}

\def \di {\partial}

\def \Su {\mathop{\sum}\limits}
\def \vspe {\mathop{}\limits_{}^{}}

\def \l. {\left.}
\def \r. {\right.}
\def \la {\left\langle}
\def \ra {\right\rangle}
\def \l| {\! \left| \,}
\def \r| {\right|}
\def \La {
\left\langle \!\!{\,}^{\mathop{}\limits_{}}_{\mathop{}\limits^{}}\right.}
\def \Ra {
\left. \!\!{\,}^{\mathop{}\limits_{}}_{\mathop{}\limits^{}}\right\rangle}
\def \Vl {
\left. \!\!{\,}^{\mathop{}\limits_{}}_{\mathop{}\limits^{}} \right|}

\def \dz {\mathfrak{z}}
\def \Ll {\mathcal{L}}

\begin{document}

\title{Global Conformal Invariance and Bilocal Fields with
	Rational Correlation Functions\footnote{Invited talk,
  presented by I.T. Todorov, at the Third International
  Sakharov Conference on Physics, Moscow, June 24--29, 2002;
  to be published in the Proceedings}}

\authors{%
 N.{\,}M. Nikolov,\adref{1}\footnote{mitov@inrne.bas.bg}
 Ya.{\,}S. Stanev,\adref{1,2}\footnote{Yassen.Stanev@roma2.infn.it} and
 I.{\,}T. Todorov\,\adref{1,3}\footnote{todorov@inrne.bas.bg}}

\addresses{\1ad Institute for Nuclear Research and Nuclear Energy,
  Tsarigradsko Chaussee 72, BG--1784 Sofia, Bulgaria,
  \nextaddress
  \2ad Dipartimento di Fisica, Universit\`{a} di Roma ``Tor Vergata'',
  I.N.F.N.{\ }--{\ }Sezione di Roma{\,}II,
  Via della Ricerca Scientifica 1, I--00133, Roma, Italy,
  \nextaddress
  \3ad Erwin Schr\"{o}dinger International Institute for
  Mathematical Physics,
  Boltzmanngasse 9, A--1090 Wien, Austria}

\maketitle

\begin{abstract}
	The singular part of the \textit{operator product expansion} (OPE)
	of a pair of \textit{globally conformal invariant} (GCI) scalar fields
	$\phi$ of (integer) dimension $d$ can be written as a sum of the
	$2$--point function of $\phi$ and $d-1$ bilocal conformal fields
	$V_{\nu} \left( x_1,\, x_2 \right)$ of dimension
	$\left(  \nu,\, \nu\right)\,$, \(\nu  = 1,\, \dots ,\, d-1\,\).
	As the correlation functions of $\phi \left( x \right)$ are
	proven to be rational \cite{NT 01}, we argue that
	the correlation functions of $V_{\nu}$ can also
	be assumed rational.
	Each \(V_{\nu} \left( x_1,\, x_2 \right)\) is
	expanded into local symmetric tensor fields
	of \textit{twist} (dimension minus rank) $2\nu\,$.
	The case \(d=2\,\), considered previously \cite{NST 02}, is briefly
	reviewed and current work on the \(d=4\) case (of a Lagrangean
	density in $4$ space--time dimensions) is previewed.
\end{abstract}

\vspace{13pt}


\vspace{0.1in}


\vspace{0.15in}

\section{Introduction}

Our study \cite{NST 02} of the theory of a GCI hermitean scalar field of
dimension $2$ suggests the following generalization.

Given a GCI neutral scalar field $\phi$ of dimension
\(d \left( \in \, \N \right)\) in $4$ dimensional Minkowski space
we look for an OPE of the product of two $\phi$'s in
\textit{bilocal fields}:
\beq\label{1.1}
\phi \left( x_1 \right) \, \phi \left( x_2 \right) \, = \,
\la 12 \,\ra \, + \,
\Su_{\nu \, = \, 1}^{d-1} \, \left( 12 \right)^{d-\nu}
\, V_{\nu} \left( x_1,\, x_2 \right) \, + \,
: \! \phi \left( x_1 \right) \phi \left( x_2 \right) \! :
\, . \qquad
\eeq
Here we are using the following shorthand notation of
\cite{NST 02}:
\beqa\label{1.2}
&&
\la 1 \dots  n \,\ra \, = \,
\lvac \phi \left( x_1 \right) \dots \,\phi \left( x_n \right) \,\rvac
\, , \quad
\nn &&
\la 12 \,\ra
\, = \, N_{\phi} \left( 12 \right)^{d}
\, , \quad
\left( 12 \,\right) \, = \, \frac{1}{4 \pi^2 \rho_{12}}
\, , \quad
\rho_{12} \, = \, x_{12}^{\ 2} + i \, 0 \, x_{12}^0
\eeqa
(the metric signature is space--like: \(x^2 := \mathbf{x}^2 - x_0^2\,\),
\(\mathbf{x}^2 = x_1^2 + x_2^2 + x_3^2\,\).)
The bilocal conformal field \(V_{\nu} \left( x_2,\, x_2 \right)\)
of dimension $\left( \nu,\, \nu \right)$ can be expanded in a
series of twist $2\nu$ local symmetric traceless tensor
fields.
Each term in this expansion is universal, only the (numerical)
coefficients depend on the theory{\ }--{\ }i.{\,}e., on the
dimension $d$ and (possible additional assumptions on) the field
$\phi\,$.
As a consequence, the fields $V_{\nu}$ and the \textit{normal product}
\(:\! \phi \left( x_1 \right) \phi \left( x_2 \right) \! :\)
defined by (\ref{1.1}) are mutually orthogonal:
\beqa\label{1.3}
&&
\lvac V_{\nu} \left( x_1,\, x_2 \right) \,\rvac \, = \, 0
\, = \, \lvac V_{\lambda} \left( x_1,\, x_2 \right) \,
V_{\nu} \left( x_3,\, x_4 \right) \,\rvac
\, , \ \text{for} \quad
\lambda \, \neq \, \nu
\, , \quad
\nn &&
\lvac V_{\nu} \left( x_1,\, x_2 \right)
:\! \phi \left( x_3 \right) \phi \left( x_4 \right) \! :\,\rvac \, = \, 0
\, , \quad \lambda,\, \nu \, = \, 1,\, \dots,\, d-1
\, . \qquad
\eeqa

Let $s_{ij}$ be the substitution exchanging the arguments $x_i$ and $x_j$
of a function of several $4$--vectors.
Then the symmetrized contribution $F_{\nu}$ of twist $2\nu$ tensor fields
to the truncated $4$--point function:
\beqa\label{1.4}
\W_4^t \left( d \right) \, \equiv && \!\!\!\!\!\!
\Wt \left( x_1,\, x_2,\, x_3,\, x_4;\, d \right) \, := \,
\nn = && \!\!\!\!\!\!
\la 1234 \,\ra - \la 12 \,\ra \la 34 \,\ra -
\la 13 \,\ra \la 24 \,\ra - \la 14 \, \ra \la23 \,\ra
\, , \quad
\eeqa\beqa\label{1.5}
F_{\nu} \left( x_{12},\, x_{23},\, x_{34};\, d \right) \, = && \!\!\!\!\!\!
\left( 1 + s_{23} + s_{13} \right) \left( 12 \right)^{d-\nu}
\left( 34 \right)^{d-\nu}
\, \text{{\small \(\times\)}}
\nn  && \!\!\!\!\!\!\ \text{{\small \(\times\)}} \,
\lvac V_{\nu} \left( x_1,\, x_2 \right) \,
V_{\nu} \left( x_3,\,  x_4 \right) \, \rvac
\, , \qquad
\eeqa
is a crossing symmetric rational function of $\rho_{ij}$ (\ref{1.2}).
The vacuum expectation value of the product of two $V_1$ can
be written as
\beqa\label{1.6}
\lvac V_1 \left( x_1,\, x_2 \right) \,
V_1 \left( x_3,\, x_4 \right) \, \rvac \, = && \!\!\!\!\!\!
\left[ \left( 13 \right) \left( 24 \right) +
\left( 14 \right) \left( 23 \right) \right]
f \left( s,\, t \right) \, = \,
\nn = && \!\!\!\!\!\!
\left( 13 \right) \left( 24 \right) \left( 1+t^{-1} \right)
f \left( s,\, t \right)
\, , \qquad
\eeqa
where \(f  \left( = f_{d} \right)\) is a function of the
conformally invariant cross ratios
\beq\label{1.7}
s \, = \, \frac{\rho_{12}\rho_{34}}{\rho_{13}\rho_{24}}
\, , \quad
t \, = \, \frac{\rho_{14}\rho_{23}}{\rho_{13}\rho_{24}}
\, , \qquad
\eeq
satisfying the $s_{12}$--symmetry condition
\beq\label{1.8}
\left( s_{12} f \right) \left( s,\, t \right)
\, := \, f \left(\!\vspe\right. \frac{s}{t},\, \frac{1}{t} \left.\!\vspe\right)
\, = \,
f \left( s,\, t \right)
\, . \qquad
\eeq

The present paper is devoted to a general study of the field
$V_1 \left( x_1,\, x_2 \right)$ which involves the (even  rank)
conserved symmetric traceless tensors
\beq\label{1.9}
T_l \left( x,\, \zeta \right) \, = \,
T\left( x \right)_{\mu_1\, \dots \, \mu_l} \,
\zeta^{\mu_{1}} \!\dots\,\zeta^{\mu_{l}}
\, , \quad
\Box_{\zeta} \, T_l \left( x,\, \zeta \right) \, = \,
0 \, = \, \frac{\di^{\, 2}}{\di x^{\mu} \,\di \zeta_{\mu}} \,
T_l \left( x,\, \zeta \right)
\,
\eeq
in its expansion in local fields.
The simplification in this case stems from the fact that all
$3$--point functions
\(\lvac V_1 \left( x_1,\, x_2 \right)\,
T_{2l} \left( x_3,\, \zeta \right)\,\rvac\,\),
whose expressions are derived from conformal invariance alone,
satisfy the d'Alembert equation in both $x_1$ and $x_2\,$; as a
result, so does $V_1\,$:
\beq\label{1.10}
\Box_1 \, V_1 \left( x_1,\, x_2 \right) \, = \, 0 \, = \,
\Box_2 \, V_1 \left( x_1,\, x_2 \right)
\, , \quad
\Box_j \, = \, \frac{\di^{\, 2}}{\di x_j^{\mu}\,\di x_{j\,\mu}}
\, , \quad j \, = \, 1,\, 2 \, . \qquad
\eeq
This allows to  compute the function $f_d \left( s,\, t \right)$
(of Eq. (\ref{1.6}))
in terms of \(d-1\) (real)
constants.
For \(d=2\) the commutator algebra generated by
$V_1 \left( x_1,\, x_2 \right)$ is relatively simple: it coincides
with a central extension of the infinite (real) symmetric Lie algebra
$\widehat{\mathit{sp}} \left( \infty \right)$ \cite{NST 02}.

GCI allows to formulate the theory in compactified Minkowski space
$\M\,$, which admits a convenient realization as a $U \left( 1 \right)$
bundle over $\Sr^3\,$:
\beqa\label{1.11}
\M \, = \, \frac{\Sr^3 \times \Sr^1}{\Z_2} \, = \,
\left\{\!\vspe\right.
z_{\alpha} \, = \,e^{2\pi i \vartheta} \, u_{\alpha} \, ;
&& \!\!\!\!\!
\vartheta \, \in \, \Fraa{\R}{\Z}\, ,\ \left( u_{\mu} \right) \, = \,
\left( \mathbf{u},\, u_4 \right) \, \in \, \Sr^3
\nn  && \!\!\!\!\!
\text{(i.e. \(u^2 \, = \, \mathbf{u}^2 + u_4^2 \, = \, 1\))}
\left.\vspe\!\right\}
\, , \qquad
\eeqa
single valued (observable) fields being periodic of period $1$
with respect to the conformal time variable $\vartheta$
(moreover, $z_{\alpha}$ does not change for
\(\vartheta \mapsto \vartheta + \frac{\textstyle 1}{\textstyle 2}\,\),
\(u_{\alpha} \mapsto -\, u_{\alpha}\,\)).
The passage from the real Minkowski space coordinates $x_{\mu}$ to
the complex $\M$ coordinates $z_{\alpha}$ is given by the complex
conformal transformation
\beq\label{1.12}
\mathbf{z} \, = \, \frac{\mathbf{x}}{\omega \left( x \right)}
\, , \ \,
z_4 \, = \, \frac{1-x^{\, 2}}{2\,\omega \left( x \right)}
\, , \ \,
\omega \left( x \right) \, = \, \frac{1+x^{\, 2}}{2} \, - \, i \, x^0
\ \,
(\,
z^{\, 2} \, := \, \mathbf{z}^{\, 2} + z_4^2 \, = \,
\frac{\overline{\omega \left( x \right)}}{\omega \left( x \right)}
\, )
\,
\eeq
and by an accompanying field
transformation
(see \cite{Tod 86} or Sec. 4 of \cite{NST 02}).
The compact picture fields have a natural decomposition  into
discrete modes.
The presence of such a discrete basis simplifies  the study of the
unitarity condition for the vacuum representation of
$\widehat{\mathit{sp}} \left( \infty \right)$ \cite{NST 02}
reviewed in Sec. 2.

Sec. 3 is devoted to the study of the contribution of $V_1$
to the truncated $4$--point function $\Wt_4$ in the case of a
field $\Ll$ of dimension $4$ with the properties of a
(gauge invariant) Lagrangean density.
We demonstrate that this contribution can only be recovered by
the Lagrangean $\Ll_0$ of a \textit{free} (Maxwell) field for a
special choice (\(a_1 = a_2\)) of the two parameters involved
(after having excluded the presence of a \(d=2\) scalar field).
In Sec. 4 we construct the crossing symmetrized contribution
of the twist $4$ tensor fields to $\Wt_4$ which involves two more
parameters.
These results indicate that
there is room for a nontrivial GCI theory of a \(d=4\) field with rational
correlation functions.

\section{OPE for $V_1 \left( x_1,\, x_2 \right)\,$.
Infinite symplectic algebra for \(d=2\) and Hilbert space positivity}

The story of conformally invariant OPE goes back over 30
years{\ }--{\ }see \cite{FGGP 72-73}; for a sample of later reviews
and further developments we refer to
\cite{D-T 77, TMP 78, OP 94, FP 98, DO 01}.
Here we summarize
and extend to an arbitrary $\phi$ the results of Sec. 3 and Appendix  A
of \cite{NST 02}.

For any dimension $d$ of the basic field $\phi$ the expansion
of $V_1$ into local tensor fields of type (\ref{1.9}) assumes
the form
\beq\label{2.1}
V_1 \left( x_1,\, x_2 \right) \, = \, \Su_{l \, = \, 0}^{\infty} \,
C_{l} \left( d \right) \,
\mathop{\int}\limits_{\!\!\!\!\!\!\! 0}^{\,\,\,\,\, 1} \, \mathit{d} \alpha
\, K_l \left( \alpha,\, \rho_{12}\, \Box_2 \right) \,
T_{2l} \left( x_2+\alpha x_{12},\, x_{12} \right)
\, , \qquad
\eeq
where
\beq\label{2.2}
K_l \left( \alpha,\, z \right) \, = \, \frac{\left( 4\, l + 1 \right)!}{
\left[ \left( 2\, l \right)! \right]^2} \
\alpha^{2l} \, \left( 1-\alpha \right)^{2l} \, \Su_{n \, = \, 0}^{\infty} \,
\frac{\left[ \alpha\left( \alpha-1 \right)
\frac{\textstyle z}{\textstyle 4} \right]^n}{
n! \, \left( 4l+1 \right)_n}
\, , \quad
\left( \nu \right)_n \, = \, \frac{\Gamma \left( \nu+n \right)}{
	\Gamma \left( \nu \right)}
	\, ,
\eeq
$\Box_2$ is the d'Alembert operator acting on $x_2$ for fixed
$x_{12}\,$.
The  integro-differen{\-}ti{\-}al operator in (\ref{2.1}) transforms
the $2$--point function of $T_{2l}$ into a $3$--point one
(see  \cite{DO 01}); for light--like auxiliary variable
$\zeta$ this relation assumes the form:
\beq\label{2.3}
\mathop{\int}\limits_{\!\!\!\!\!\!\! 0}^{\,\,\,\,\, 1} \!\! \mathit{d} \alpha
\, K_l \left( \alpha,\, \rho_{12}\, \Box_2 \right) \,
\frac{\left( x_{12} \!\cdot\! r \left( y \left( \alpha \right) \right)
\!\cdot\! \zeta \right)^{2l}}{
\rho_{y \left( \alpha \right)}^{2l+2}} \, = \,
\frac{\left( X \!\cdot\! \zeta \right)^{2l}}{\rho_{13} \, \rho_{23}}
\, , \ \,
\text{with} \
y \left( \alpha \right) \, := \, x_{23} + \alpha \, x_{12}
\, ; \
\eeq
here
\beqa\label{2.4}
&& \!\!\!\!\!\!\!\!\!\!
X \, = \, X_{12}^3 \, := \, \frac{x_{13}}{\rho_{13}} \, - \,
\frac{x_{23}}{\rho_{23}}
\, , \quad
\zeta_1 \cdot r \left( y \right)
\cdot \zeta_2 \, = \, \zeta_1 \cdot \zeta_2 \, - \, 2 \,
\frac{\left( \zeta_1 \!\cdot\! y \right)\left( \zeta_2 \!\cdot\! y \right)}{
\rho_y}
\, , \qquad
\\ \label{2.5}
&& \!\!\!\!\!\!\!\!\!\!
\rho_{y \left( \alpha \right)} \, = \, y^2 \left( \alpha \right) \, + \,
i \, 0 \, y \left( \alpha \right)^0 \, = \,
\alpha \, \rho_{13} \, + \, \left( 1-\alpha \right) \rho_{23} \, - \,
\alpha \left( 1-\alpha \right)\rho_{12}
\, . \qquad
\eeqa
(Had we started with a complex scalar field $\phi$ and
with the {\small OPE} of
\(\phi^* \left( x_1 \right) \phi \left( x_2 \right)\)
instead of (\ref{1.1}) we would have also encountered odd rank symmetric tensors
\(T_{2l+1} \left(  x,\, x_{12} \right)\) in the expansion of
\(V_1 \left( x_1,\, x_2 \right)\,\).)

The contribution of $T_{2l}$ to the $4$--point function (\ref{1.6})
is universal (up to normalization) and is expressed in terms of
hypergeometric functions
(see Eq. (3.10) of \cite{DO 01} or (A.6) of \cite{NST 02}).
It is determined by its value on the light cone
\(\rho_{34} \, = \, 0 \, ( \, = \, s\, )\) (cf. Appendix A of \cite{NST 02})
for which its expression is particularly simple:
\beqa\label{2.6}
& \lvac \, V_1 \left( x_1,\, x_2 \right) \,
\mathop{\text{{\Large \(\int\)}}}\limits_{\!\!\!\! 0}^{\,\, 1} \!
\mathit{d} \alpha
\,
\text{{\large \(\frac{\alpha^{2l} \left( 1-\alpha \right)^{2l}}{
B \left( 2l+1,\, 2l+1 \right)}\)}}
\
T_{2l} \left( x_4+\alpha x_{34},\, x_{34} \right) \, \rvac \, = &
\nn &
= \,
A_l \, \left( 13 \right) \left( 24 \right) \left( 1-t \right)^{2l} \,
F \left( 2l+1,\, 2l+1;\, 4l+2;\, 1-t \right)
& \nn &
(\, B \left( \mu,\, \nu \right)
\, := \, \mathop{\text{{\Large \(\int\)}}}\limits_{\!\!\!\! 0}^{\,\, 1} \!
x^{\mu-1} \left( 1-x \right)^{\nu-1} \, \mathit{d} x
\, = \,
\text{{\large \(\frac{\Gamma \left( \mu \right) \, \Gamma \left( \nu \right)}{
\Gamma \left( \mu+\nu \right)}\)}}
\, )
\, . &
\eeqa
The constant $A_l$ (proportional to $C_l \left( d \right)$) provides
the normalization of the $3$--point function
\beqa\label{2.7}
&
\lvac \, V_1 \left( x_1,\, x_2 \right) \, T_{2l} \left( x_3,\, \zeta \right) \,
\rvac \, = \, A_l \left( 12 \right) \left( X^{\, 2} \right)^{l+1}
\left( \zeta^{\, 2} \right)^l \,
C_{2l}^1 \left( \widehat{\zeta} \!\cdot\! \widehat{X} \right)
& \nn &
(\, = \,
\text{{\large \(\frac{A_l}{\left( 2 \, \pi \right)^2} \
\frac{\left( 2 \, \zeta \cdot X \right)^{2l}}{\rho_{13} \, \rho_{23}}\)}}
\ \, \text{for} \ \, \zeta^{\, 2} \, = \, 0 \, )\, ,\quad
\widehat{v} \, = \,
\text{{\large \(\frac{v}{\sqrt{v^{\, 2\,}}^{{}_{}^{}}_{{}_{}^{}}}\)}}
\quad
(\,A_l \, = \, N_l \, C_l \, )
\, \qquad
&
\eeqa
(\(N_l > 0\) fixing the normalization of the $2$--point function of $T_{2l}$).
The expansion parameter \(1-t\)
tends to zero whenever the light--like vector $x_{34}$ does:
\beq\label{2.8}
1-t \, = \, 2 \left( \frac{x_{24}}{\rho_{24}} \, - \,
\frac{x_{13}}{\rho_{13}} \right) \!\cdot x_{34}
\, + \, 4 \, \frac{\left( x_{13} \!\cdot x_{34}\! \right)
\left( x_{24} \!\cdot\! x_{34} \right)}{\rho_{13} \, \rho_{24}}
\quad \text{for} \quad \rho_{34} \, = \, 0
\, . \qquad
\eeq

If we change the normalization of $T_{2l}$ setting
\(T_{2l} \mapsto Z_l \, T_{2l}\) (keeping $\phi$ fixed) then the constants
$C_l \left( d \right)$ and $A_l$ appearing in (\ref{2.1}) and (\ref{2.7})
will also change,
\(A_l \mapsto Z_l \, A_l\,\), \(C_l \mapsto Z_l^{-1} \, C_l\)
(\(N_l \mapsto Z_l^2 \, N_l\)), but their product \(A_l \, C_l\)
will remain the same and it can be determined by the  $4$--point
function (\ref{1.6}).
We, therefore, proceed to compute the general form of this $4$--point
function.

According to (\ref{1.10}) it satisfy the d'Alembert equation in $x_j\,$.
This implies a second order partial differential equation for
\(f_d \left( s,\, t \right)\,\):
\beqa\label{2.9}
\left\{\!\vspe\right.
&& \!\!\!\!\!\!\!\!\!
\left( 1+t^2-s \right) D_s + s\left( t-1 \right) D_t +
\left( 1+t \right)
\left[{}_{{}_{{}_{}}}^{{}_{{}_{}}^{{}^{}}}\!\!\right.
t \, D_s^2 + s \, D_t^2 +
\nn
&& \!\!\!\!\!\!\!\!\!
\ +\, \left( s+t-1 \right) D_s \, D_t
\left.{}_{{}_{{}_{}}}^{{}_{{}_{}}^{{}^{}}}\!\!\right]
\left.\vspe\!\right\} f_d \left( s,\, t \right)
\, = \, 0
\, , \quad
D_u \, = \, u \frac{\di}{\di u}
\quad (\, u \, = \, s,\, t \, )
\, . \qquad
\eeqa

We are looking for a solution of (\ref{2.9}) for which the
product
\(t^{d-2} \left( 1+t \right)
\text{{\scriptsize $\times$}}\linebreak
f_d \left( s,\, t \right)\)
is a polynomial in $s$ and $t$ of overall degree not exceeding $2d - 3\,$.
(This requirement follows from Wightman positivity \cite{SW 64-00}:
for space--time
dimensions \(D>2\,\), the truncated $4$--point function has a strictly
smaller singularity for \(\rho_{ij} \to 0\)
(\(1 \leq i < j \leq 4\)) than the corresponding $2$--point
function{\ }--{\ }see \cite{NT 01, NST 02}.)
Such a solution can be obtained starting with an ``initial condition''
in $s$ that obeys the symmetry property (\ref{1.8}):
\beq\label{2.10}
f_d \left( 0,\, t \right) \, = \,
\Su_{\nu \, = \, 0}^{d-2} \, a_{\nu} \, \frac{\left( 1-t \right)^{2\nu}}{
t^{\nu}} \ \, (\ = \, f_d
\left(\vspe\!\right.
0,\, \frac{1}{t}
\left.\vspe\!\right)
\, )
\, . \qquad
\eeq
In particular, for \(d=2\,\), $f_2 \left( s,\, t \right)$
is a constant (which we shall denote by $c$ since it is analogous to
the Virasoro central charge{\ }--{\ }\cite{NST 02}).
In this case the products \(B_l := A_l \, C_l \left( 2 \right)\)
can be calculated from the seemingly overdetermined infinite set of
algebraic equations stemming from
\beqa\label{2.11}
c \,
\left(\!\!\vspe\right.
1+\, \frac{1}{t} \,
\left.\vspe\!\right) \ (\ = && \!\!\!\!\!\!
c \,
\left\{\!\vspe\right.
2 + \Su_{n \, = \, 1}^{\infty} \, \left( 1-t \right)^n
\left.\vspe\!\right\}
\ ) \, = \,
\nn = && \!\!\!\!\!\!
\Su_{l \, = \, 0}^{\infty} \, B_l \left( 1-t \right)^{2l}
F \left( 2l+1,\, 2l+1;\, 4l+2;\, 1-t  \right)
\, . \qquad
\eeqa
The result is (\cite{NST 02})
\beq\label{2.12}
B_l \ \, ( \, = \, A_l \, C_l \left( 2 \right) \, = \,
N_l \, C_l^2 \left( 2 \right) \, ) \, = \, \frac{2\, c}{
\text{{
\small \(\left( \!\!\!\begin{array}{c} 4 \, l \\ 2 \, l \end{array}\!\! \right)\)
}}
}
\quad
(\, \text{i.{\,}e.,} \ \,  B_0 \, = \, 2\, c\, , \ \, B_1 \, = \,\frac{c}{3}
\ \ \ \text{etc.} \, )
\, .
\eeq

We see that for \(c > 0\) all expansion coefficients (\ref{2.12})
are positive thus \textit{a necessary condition for Hilbert space}
(\textit{Wightman}) \textit{positivity} is satisfied.
This condition is not, however, sufficient.
We shall briefly review the argument of \cite{NST 02} which
establishes a necessary and sufficient positivity condition.

One first observes that
for \(c = N \, ( \in \, \N \, )\) the OPE algebra of the bilocal
field \(V_1 \left( x_1,\, x_2 \right)\) coincides
(for \(d=2\)) with the algebra of a sum of normal products of
mutually commuting free massless scalar fields $\varphi_i\,$:
\beq\label{2.13}
V_1 \left( x_1,\, x_2 \right) \, = \, \Su_{i \, = \, 1}^{N} \,
:\! \varphi_i \left( x_1 \right) \varphi_i \left( x_2 \right) \! :
\quad \text{for} \quad c \, = \, N \
(\, = \, 1,\, 2,\, \dots \, )
\, , \qquad
\eeq
where
\beq\label{2.14}
\lvac \, \varphi_i \left( x_1 \right) \varphi_j \left( x_2 \right) \,
\rvac \, = \, \delta_{ij} \, \left( 12 \,\right)
\ ( \ = \, \delta_{ij} \ \frac{1}{4\, \pi^2 \, \rho_{12}} \ )
\, . \qquad
\eeq
Since each $\varphi_i$ generates under commutation an (infinite)
Heisenberg (Lie) algebra, it follows that $V_1$ generates a central
extension of the infinite (real) symplectic algebra.

To find a stronger positivity condition for the vacuum
representation of $V_1$ we shall, following Sec. 5 of \cite{NST 02},
use the compact picture field $V_1 \left( z_1,\, z_2 \right)$
(expressed in terms of the coordinates (\ref{1.11})).
To this end we observe that in accord with a general theorem of
Borchers \cite{B 64} the restriction of $V_1$ to a $1$--dimensional bilocal
field $v \left( \dz_1,\, \dz_2 \right)$ for fixed $u$
exists and admits a discrete mode expansion of the type:
\beq\label{2.15}
v \left( \dz_{\, 1},\, \dz_{\, 2} \right)
\ (\ := \, V_1 \left( \dz_{{}_{} 1} u,\, \dz_{{}_{} 2} u \right) \ ) \, = \,
\Su_{n,\, m \, \in \, \Z} \, v_{nm} \, \dz_{{}_{}1}^{-n-1} \, \dz_{{}_{}2}^{-m-1}
\, . \qquad
\eeq
Here $v_{nm}$ satisfy the following
\(\widehat{\mathit{sp}} \left( \infty\right)\) commutation relations:
\beqa\label{2.16}
\left[ \, v_{n_1m_1} \, ,\, v_{n_2m_2} \, \right] \, = && \!\!\!\!\!\!
c \, n_1 \, m_1 \,
\left(\!\vspe\right.
\delta_{n_1,\, -\, n_2} \, \delta_{m_1,\, -\, m_2}
\, + \,
\delta_{n_1,\, -\, m_2} \, \delta_{m_1,\, -\, n_2}
\left.\!\vspe \right) \, + \,
\nn && \!\!\!\!\!\!
\, + \
n_1 \,
\left(\!\vspe\right.
\delta_{n_1,\, -\, n_2} \, v_{m_1m_2}
\, + \,
\delta_{n_1,\, -\, m_2} \, v_{m_1n_2}
\left.\!\vspe \right) \, + \,
\nn && \!\!\!\!\!\!
\, + \
m_1 \,
\left(\!\vspe\right.
\delta_{m_1,\, -\, n_2} \, v_{n_1m_2}
\, + \,
\delta_{m_1,\, -\, m_2} \, v_{n_1n_2}
\left.\!\vspe \right)
\, . \qquad
\eeqa
The vacuum representation of $\widehat{\mathit{sp}} \left( \infty \right)$
is characterized by the standard conditions
\beq\label{2.17}
v_{nm} \, \rvac \, = \, 0 \, = \, \lvac \, v_{-\, n,\, -\, m}
\quad \text{if} \quad n \, \geq \, 0 \quad \text{or} \quad m \, \geq \, 0
\, . \qquad
\eeq
Consider the vector
\beq\label{2.18}
\la {}_{}\Delta_n {}_{}\right| \, = \, \frac{1}{n!} \
\lvac \,
\left|\begin{array}{llll}
v_{11} & v_{12} & \dots & v_{1n} \\
v_{21} & v_{22} & \dots & v_{2n} \\
\dots  & \dots  & \dots & \dots  \\
v_{n1} & v_{n2} & \dots & v_{nn} \\
\end{array}\!\!\right|
\, \qquad
\eeq
which vanishes for $V$ given by (\ref{2.13}) and \(n>N\,\).
Its norm square is given by (\cite{NST 02} Lemma 2.5)
\beq\label{2.19}
\left\| {}_{\!}^{\!} \left| {}_{}\Delta_n {}_{} \ra {}_{}^{} \right\|^2
\, = \,
\La \Delta_n \!\Vl \Delta_n \!\Ra \, = \, \left( n+1 \right)! \
c \, \left( c-1 \right) \dots \left( c-n+1 \right)
\, . \qquad
\eeq

It follows that a necessary and sufficient condition
for the unitarity of the vacuum representation of
$\widehat{\mathit{sp}} \left( \infty \right)$
is \(c \in \N\) (\cite{NST 02} Theorem 5.1).
For positive integer $c\,$, however, the \(d=2\) scalar field
\(\phi \left( x \right) \, = \, \frac{\textstyle 1}{\textstyle 2^{}_{}} \
V_1 \left( x,\, x \right)\) can be presented according to (\ref{2.13})
as a sum of normal products of free fields.

This somewhat disappointing result is based on the fact that the general
GCI $4$--point function of $\phi$ (for \(d=2\)) is a multiple of the one
for the normal square of a free field.
It appears, therefore, promising to study the theory of a \(d=4\)
GCI field for which, as we are going to demonstrate, this is not the case.

\section{Lagrangean density. Expansion in twist $2$ tensors of the bilocal
field \(V_1 \left( x_1,\, x_2 \right)\)}

A GCI model of a \(d=4\) scalar field \(\Ll \left( x \right)\) in $4D$
Minkowski space is of particular interest since it may provide a
non--perturbative description of a renormalization group fixed
point of a non--abelian gauge theory (in which there is no gauge invariant
field of dimension lower than $4\,$).

We begin by writing down the solution of Eq. (\ref{2.9}) satisfying
the initial condition (\ref{2.10}) and the symmetry relation (\ref{1.8})
for \(d=4\) assuming there is no \(d=2\) scalar field in the theory
(i.{\,}e. setting \(a_0 = 0\,\)):
\beqa\label{3.1}
f_4 \left( s,\, t \right) \, = && \!\!\!\!\!\!
a_1 \,
\left\{\!{\vspe}^{\vspe}_{\vspe}\right.
\frac{\left( 1\! -\! t \right)^2}{t}
\left(
1 \, - \, \frac{s}{1\! +\! t}
\right)
\, - \,
\frac{2s}{1\! +\! t}
\left.\!{\vspe}^{\vspe}_{\vspe}\right\}_{\vspe\!}
\, + \,
\nn && \!\!\!\!\!\!
\hspace{-50pt}
\, + \,
a_2 \,
\left\{\!{\vspe}^{\vspe}_{\vspe}\right.
\frac{\left( 1\! -\! t \right)^4}{t^2}
\left(
1 \, - \, \frac{2\, s}{1\! +\! t}
\right)_{\vspe\!}
\, - \,
\frac{6 \, s}{1\! +\! t} \ \frac{\left( 1\! -\! t \right)^2}{t}
\, + \,
\frac{s^2}{t}
\left(
1 \, + \, \frac{\left( 1\! -\! t \right)^2}{t}
\right)
\left.\!{\vspe}^{\vspe}_{\vspe}\right\}^{\vspe}
\, . \qquad
\eeqa
The corresponding crossing symmetric contribution $F_1$
(\ref{1.5}), (\ref{1.6}) to the truncated $4$--point function reads:
\beqa\label{3.2}
&&
F_1 \left( x_{ij};\, d = 4 \right) \, = \,
\left( 12 \right)^2 \left( 23 \right)^2 \left( 34 \right)^2
\left( 14 \right)^2 \frac{1}{st} \ \Su_{\nu \, = \, 1}^2 \,
a_{\nu} \, I_{\nu} \left( s,\, t \right)
\, , \quad
\nn &&
s^5 \, I_{\nu} \left(\!\vspe\right.
\frac{1}{s},\, \frac{t}{s} \left.\vspe\!\right)
\, = \, I_{\nu} \left( s,\, t \right) \, = \, I_{\nu} \left( t,\, s \right)
\, , \qquad
\eeqa
where $I_{\nu}$ are polynomials in the cross ratios:
\beqa\label{3.3}
&& \!\!\!\!\!\!\!\!\!\!
I_1 \left( s,\, t \right) \, = \,
t \, \left[ \left( 1-t \right)^2 \left( 1+t \right) - \,
s \left( 1+t^2 \right)\right] + \, \!\!\!\!\!
\nn && \!\!\!\!\!\!\!\!\!\! \ \ \quad + \,
s \, \left[ \left( 1-s \right)^2 \left( 1+s \right) - \,
t \left( 1+s^2 \right)\right] + \,
s\, t \left[ \left( s-t \right)^2 \left( s+t \right) \, - \,
s^2 \, - \, t^2 \right]
\, , \qquad \vspe \!\!\!\!\!
\nn && \!\!\!\!\!\!\!\!\!\!
I_2 \left( s,\, t \right) \, = \,
\left( 1\! -t \right)^2 \left( 1\! +t \right)
\left( 2\! - 3 \, t + 2 \, t^2 \right)
+
\left( 1\! -s \right)^2 \left( 1\! +s \right)
\left( 2\! - 3 \, s + 2 \, s^2 \right) - \, \vspe  \!\!\!\!\! \!\!\!\!\!
\nn && \!\!\!\!\!\!\!\!\!\! \ \ \quad - \, 2 \, + \, 4 s \, t
\left( s^2 + t^2 +1 \right)
-
s \, t
\left( s\! +\! t \right)
\left( 5\, s^2 - 8 \, s\, t +5\, t^2 \right)
\, . \qquad \vspe \!\!\!\!\!
\eeqa

We shall demonstrate that the expression (\ref{1.6})
for $f_4$ given by (\ref{3.1}) is reproduced by an
expansion of $V_1$ in conserved symmetric traceless tensors.
In view of (\ref{2.1}) and (\ref{2.6}) this
amounts to determining the (invariant under rescaling)
structure constants
\(B_l \, ( \, = N_l \, C_l^2 \left( 4 \right))\) in such a way that
\beqa\label{3.4}
&&
\Su_{l \, = \, 1}^{\infty} \, B_l \left( 1-t \right)^{2l}
F \left( 2l+1,\, 2l+1;\, 4l+2;\, 1-t \right) \, = \,
\nn && \quad
\, = \,
\left( 1+\frac{1}{t} \right) \left\{\!{\vspe}_{\vspe}^{\vspe}\right.
a_1 \, \frac{\left( 1\! -t \right)^2}{t} \ + \
a_2 \, \frac{\left( 1\! -t \right)^4}{t^2} \left.{\vspe}_{\vspe}^{\vspe}\!\right\}
\, . \qquad
\eeqa
We note that the system (\ref{3.4}) is overdetermined:
each $B_l$ must satisfy two conditions to fit the coefficients
to $\left( 1 -t\right)^{2l}$ and $\left( 1-t \right)^{2l+1}\,$.
Thus, the existence of a solution provides a non--trivial
consistency check.
One verifies that such a solution does exist and is given by
\beq\label{3.5}
B_l \, = \, \frac{\left( 2l \right)! \, \left( 2l\! +\! 1 \right)!}{
\left( 4l\! -\! 1 \right)!} \ \left[ \, a_1 + \,
\frac{\left( 2l\! +\! 3 \right)\left( l\! -\! 1 \right)}{2} \ a_2 \, \right]
\, . \qquad
\eeq
It is consistent with Wightman positivity for \(a_1 \geq 0\,\),
\( a_2 \geq 0\,\).

The vanishing of $f_4 \left( s,\, t \right)$ (\ref{3.1}){\ }--{\ }and
hence of
\(\lvac \, V_1 \left( x_1,\, x_2 \right) V_1 \left( x_3,\, x_4 \right) \,
\rvac = \left( 14 \right) \left( 23 \right) \left( 1+t \right)
f_4 \left( s,\, t \right)\) (\ref{1.6}){\ }--{\ }for \(s=0\), \(t=1\)
(i.e. for \(x_{34} = 0\,\), or \(x_{12} = 0\){\ }--{\ }according
to (\ref{2.8})) implies
\beq\label{3.6}
V_1 \left( x_1,\, x_2 \right)
\, = \, x_{12}^{\mu} \, x_{12}^{\nu} \, T_{\mu\nu} \left( x_1,\, x_2 \right)
\, \qquad
\eeq
where the limit $T_{\mu\nu} \left( x,\, x \right)$ exists and is a multiple
of the stress--energy tensor (i.e., its $3$--point function with $V_1$
satisfies the Ward--Takahashi identities).

We shall now demonstrate that a normal product $\Ll$ of free fields
can only reproduce an expression of the type (\ref{3.1}--\ref{3.5}))
if it is a linear combination of free electromagnetic Lagrangeans
\beq\label{3.7}
\Ll_0 \left( x \right) \, = \,
- \ \frac{1}{4} \ :\! F_{\mu\nu} \left( x \right)
F^{\mu\nu} \left( x \right) \! :
\quad \ \text{implying} \quad \ a_1 \, = \, a_2
\, . \qquad
\eeq
In other words, only a $1$--parameter subset of the $2$--parameter
family of expressions (\ref{3.5}) belongs to the Borchers' class of
free fields.

As a first step we shall verify the implication of (\ref{3.7}) for a free
Maxwell field $F$ with $2$--point function
\beqa\label{3.8}
\lvac \, F_{\mu_1\nu_1} \left( x_1 \right) F_{\mu_2\nu_2} \left( x_2 \right) \,
\rvac =
&& \!\!\!\!\!\!\!\!
4 \, D_{\mu_1\nu_1\mu_2\nu_2} \left( x_{12} \right) =
\left\{\!\vspe\right.
\di_{\mu_1} \left( \di_{\mu_2} \, \eta_{\,\nu_1\nu_2} \! -
\di_{\nu_2} \, \eta_{\,\nu_1\mu_2} \right)
-
\nn && \!\!\!\!\!\!\!\!
- \
\di_{\nu_1} \left( \di_{\mu_2} \, \eta_{\mu_1\nu_2} \! -
\di_{\nu_2} \, \eta_{\mu_1\mu_2} \right)
\left.\!\vspe\right\}_{\vspe} \left( 12 \right)
\nn \text{i.{\,}e.} \quad \
4\,\pi^2 \, D_{\mu_1\nu_1\mu_2\nu_2} \left( x \right) \, =
&& \!\!\!\!\!\!\!\!
R_{\mu_1\mu_2} \left( x \right) \, R_{\nu_1\nu_2} \left( x \right) \, - \,
R_{\mu_1\nu_2} \left( x \right) \, R_{\nu_1\mu_2} \left( x \right)
\, \qquad \vspe
\eeqa
where
\beq\label{3.9}
R^{\mu}_{\ \,\nu} \left( x \right) \, = \,
\frac{r^{\mu}_{\ \,\nu} \left( x \right)}{ \rho_x} \, = \,
\rho_x^{-2}
\left( \rho_x \, \delta^{\,\mu}_{\ \,\nu} - 2 \, x^{\mu} \, x_{\nu} \right)
\, , \quad
\rho_x \, = \, x^{\, 2} + i \, 0 \, x^0
\, . \qquad
\eeq
The OPE of the product of two $\Ll_0$'s then has the form
\beqa\label{3.10}
\Ll_0 \left( x_1 \right) \, \Ll_0 \left( x_2 \right) \, = && \!\!\!\!\!\!\!
\la 12 \,\ra_0 \, + \ D_{\mu_1\nu_1\mu_2\nu_2} \left( x_{12} \right) \,
:\! F^{\mu_1\nu_1} \left( x_1 \right) F^{\mu_2\nu_2} \left( x_2 \right) \! :
\, + \,
\nn && \!\!\!\!\!\!\!
+ \, :\! \Ll_0 \left( x_1 \right) \Ll_0 \left( x_2 \right) \! :
\, \qquad
\eeqa
where
\beqa\label{3.11}
&&
\la 12 \,\ra_0 \, = \, \frac{3}{\left( \pi\,\rho_{12} \right)^4}
\, , \quad \
\nn &&
D_{\mu_1\nu_1\mu_2\nu_2} \left( x_{12} \right) \,
:\! F^{\mu_1\nu_1} \left( x_1 \right) F^{\mu_2\nu_2} \left( x_2 \right) \! :
\ \, = \, \frac{8}{\pi^2\,\rho_{12}^3} \ V_1 \left( x_1,\, x_2 \right)
\, \qquad
\eeqa
for $V_1$ given by (\ref{3.6}) with the bilocal tensor field
\beqa\label{3.12}
T \left( x_1,\, x_2;\, \zeta \right) \, := && \!\!\!\!\!\!\!
\zeta_{\mu} \, T^{\mu}_{\ \,\nu} \left( x_1,\, x_2 \right) \,
\zeta^{\nu} \, = \,
\nn = && \!\!\!\!\!\!\!
\frac{1}{4} \
:\! F^{\sigma\tau} \left( x_1 \right) F_{\sigma\tau} \left( x_2 \right) \! :
\zeta^{\, 2} \, - \, \zeta_{\mu} \, :\!
F^{\sigma\mu} \left( x_1 \right) F_{\sigma\nu} \left( x_2 \right) \! : \zeta^{\nu}
\, . \qquad
\eeqa
The $4$--point function of \(T\) is computed from (\ref{3.8});
setting \(\zeta^{\ 2} = 0 = \zeta'^{\ 2}\) we find
\beqa\label{3.13}
&&
\pi^4 \, \lvac \,
T \left( x_1,\, x_2;\, \zeta \right) T \left( x_3,\, x_4;\, \zeta' \right) \,
\rvac \, = \,
\left\{\!\vspe\right. R^{\,\sigma}_{\ \,\tau} \left( x_{14} \right)
\left( \zeta \!\cdot\! R \left( x_{14} \right) \!\cdot\! \zeta' \right) -
\nn && \qquad -\,
( R^{\,\sigma} \left( x_{14} \right) \!\cdot\! \zeta' )
\left( \zeta \!\cdot\! R_{\ \tau} \left( x_{14} \right) \right)
\left.\!\vspe\right\}
\left\{\!\vspe\right. R^{\,\tau}_{\ \,\sigma} \left( x_{23} \right)
\left( \zeta \!\cdot\! R \left( x_{23} \right) \!\cdot\! \zeta' \right) -
\nn && \qquad -\,
\left( \zeta \!\cdot\! R_{\ \sigma} \left( x_{23} \right) \right)
( R^{\,\tau} \left( x_{23} \right) \!\cdot\! \zeta' )
\left.\!\vspe\right\} \, + \, \left\{ \, 3 \, \leftrightarrow \, 4 \, \right\}
\, . \qquad
\eeqa
The $4$--point function of $V_1$ can be obtained from here by applying
to the result the operator
\beq\label{3.14}
\frac{1}{4} \
\left\{\!\vspe\right.
\left(\!\vspe\right.
2 \, x_{12} \cdot \frac{\di}{\di\zeta}
\left.\!\vspe\right)^2
- x_{12}^{\ 2} \, \Box_{\zeta} \,
\left.\!\vspe\right\}
\left\{\!\vspe\right.
\left(\!\vspe\right.
2 \, x_{34} \cdot \frac{\di}{\di\zeta'}
\left.\!\vspe\right)^2 -
x_{34}^{\ 2} \, \Box_{\zeta'}
\left.\!\vspe\right\}
\, \qquad
\eeq
thus recovering (\ref{1.6}), (\ref{3.1}) with \(a_1 = a_2\,\),
as stated.

Given that the expression (\ref{3.1}) (the condition \(a_0 = 0\))
excludes the presence of a \(d=2\) scalar field
in the OPE of type (\ref{3.10}), we should like to exclude
another conceivable candidate for a free theory in disguise
given by a nonabelian (\(\mathit{SU} \left( 2 \right)\))
$F_{\mu\nu}^a$ that is a normal product
of free longitudinal vector fields:
\beq\label{3.15}
F_{\mu\nu}^a \left( x \right) \, = \, - \, g \,
\epsilon^{abc} \, A^b_{\mu} \left( x \right) \,
A^c_{\nu} \left( x \right) \quad \ \text{for} \quad \
\di_{\mu} \, A^b_{\nu} \left( x \right) \, = \,
\di_{\nu} \, A^b_{\mu} \left( x \right)
\, \qquad
\eeq
\(a,\, b,\, c \, = \, 1,\, 2,\, 3\,\).
Here $A^b_{\mu}$ are (generalized) free fields with conformally
invariant $2$--point functions
\beq\label{3.16}
\lvac \,
A^a_{\mu} \left( x \right) \,
A^b_{\nu} \left( x \right) \, \rvac \, = \,
\frac{1}{2} \ \delta^{ab} \, R_{\mu\nu} \left( x_{12} \right)
\, , \qquad
\eeq
\(\epsilon^{abc}\) is the fully antisymmetric Levi--Civita tensor
(\(\epsilon^{123}=1\)).
Although the expression (\ref{3.15}) is reminiscent to a
pure gauge neither $F_{\mu\nu}^a$ nor the gauge invariant ``Lagrangean''
\beqa\label{3.17}
\Ll_g \left( x \right) \, = && \!\!\!\!\!\!
\frac{1}{4} \
:\! F_{\mu\nu}^a \left( x \right) F^{\mu\nu}_a \left( x \right) \! :
\ \, = \,
\nn = && \!\!\!\!\!\!
\frac{g^2}{4} \ :\!
\left\{\!\vspe\right. \left(
A_{\mu}^a  \left( x \right)  \, A_a^{\mu} \left( x \right) \right)^2
-
A_{\mu}^a \left( x \right) \, A^{\mu}_b \left( x \right) \,
A^b_{\nu} \left( x \right) \, A^{\nu}_a \left( x \right)
\left.\!\vspe\right\} \! :
\, \qquad
\eeqa
vanishes.
A tedious but straightforward calculation gives the following
leading term in the OPE of two $\Ll_g\,$:
\beq\label{3.18}
\Ll_{g} \left( x_1 \right) \, \Ll_g \left( x_2 \right) \, = \,
\frac{9}{4} \ g^4 \, \left( 12\, \right)^3 \,
A^a_{\mu} \left( x_1 \right) \, r^{\mu\nu} \left( x_{12} \right)
\, A^a_{\nu} \left( x_{2} \right) \, + \, O \left( 12 \right)^2
\, . \qquad
\eeq
The first term in the right hand side gives rise to both the
(standard) $2$--point function of $\Ll$ and to a
$V_1 \left( x_1,\, x_2 \right)$ for which the \(d=2\) diagonal limit
$V_1 \left( x,\, x \right)$ is non--zero contrary to our assumption.
Moreover, considering a general linear combination of the two \(d=4\)
scalars made out of a triplet of \(d=1\) vector fields
(appearing in the right hand side of (\ref{3.17})),
\beq\label{3.19}
\Ll_{\xi\eta} \left( x \right) \, = \, \xi
\left( A^a_{\mu} \left( x \right) \, A^{\mu}_a  \left( x \right) \right)^2
\, - \, \eta
\left( A_{\mu}^a \left( x \right) \, A^{\mu}_b \left( x \right) \,
A^b_{\nu} \left( x \right) \, A^{\nu}_a \left( x \right) \right)
\, \qquad
\eeq
we deduce that the leading term in the OPE,
\beq\label{3.20}
\Ll_{\xi\eta} \left( x_1 \right) \, \Ll_{\xi\eta} \left( x_2 \right)
\approx
4 \left( 14 \xi^2 \! - 16 \, \xi\, \eta \! + 11 \, \eta^2 \right)
\left( 12\, \right)^3
\left( A_{\mu}^a \left( x_1 \right) \, r^{\mu\nu} \left( x_{12} \right)
\, A_{\nu}^a \left( x_2 \right) \right)
,
\eeq
never vanishes for real \(\xi,\, \eta\,\).

\section{Twist 4 contribution. Concluding remarks}

The general GCI and crossing symmetric truncated $4$--point function
of $\Ll \left( x \right)$ depends on $5$ parameters (\cite{NST 02}).
Three of them ($a_{\nu}$) appear in the crossing symmetrized
contribution $F_1$ of twist $2$ fields.
(We excluded one of them{\ }--{\ }setting \(a_0 = 0\){\ }--{\ }by
demanding that no \(d=2\) scalar field contributes to the OPE of
two $\Ll$'s.)
The remaining two parameters appear in the general expression
for $F_2$ (defined by (\ref{1.5}) for \(\nu=2,\, d=4\,\)):
\beq\label{4.1}
F_2 \left( x_{12},\, x_{23},\, x_{34};\, 4 \right) =
\left( 12\, \right)^2 \! \left( 23\, \right)^2 \!
\left( 34\, \right)^2 \! \left( 14\, \right)^2
\left\{\!\vspe\right.
b_1 \left( 1\! + s^2\! + t^2 \right) + \,
b_2 \left( s\! + t\! + s\, t \right)
\left.\!\!\vspe\right\}
\eeq
(the symmetrized contribution of twist $4$ fields).
For \(2\, b_1 + b_2 \neq 0\) we deduce that the
OPE  of \(V_2 \left( x_1,\, x_2 \right)\) involves
$\Ll\,$, so that $\Ll$ may have a non--zero $3$--point function.

To summarize: we have constructed a $4$--parameter family of
truncated $4$--point functions $\Wt_4$ of a \(d=4\) scalar
field $\Ll \left( x \right)\,$.
Moreover, no lower dimensional fields contribute to $\Wt_4\,$.
Only a $1$--parameter subset of this $4$--parameter family
corresponds to a free (abelian gauge) field theory.

As a by-product we prove that the bilocal field $V_1$ appearing
in the OPE (\ref{1.1}) of two scalar fields is the sum of an
infinite series of twist two conserved symmetric traceless tensors.

The result can be viewed as a first step to a non-perturbative
construction of (gauge and) GCI invariant correlation function in a
non-abelian gauge theory.

\vspace{0.2in}

\noindent
\textbf{Acknowledgments.}
N.N. and I.T acknowledge partial support by the Bulgarian
National Council for Scientific Research under contract F-828.
The research of Ya. S. was supported in part by
I.N.F.N., by the EC contract HPRN-CT-2000-00122, by the EC contract
HPRN-CT-2000-00148, by the INTAS contract 99-0-590 and by the MURST-COFIN
contract 2001-025492.
All three authors acknowledge
partial support by a {\small NATO}{\,}linkage{\,}grant{\,}{\small PST.CLG.978785}.
I.T. thanks the organizers of the Third Sakharov Conference for
their hospitality in Moscow.

\end{document}